\documentclass[11pt,twoside]{article}

\usepackage{asp2014}

\aspSuppressVolSlug
\resetcounters

\begin{document}

\title{Astrophysics Source Code Library: Here we grow again!}

% Note the position of the comma between the author name and the 
% affiliation number.
% Author names should be separated by commas.
% The final author should be preceded by "and".
% Affiliations should not be repeated across multiple \affil commands. If several
% authors share an affiliation this should be in a single \affil which can then
% be referenced for several author names.
% See ManuscriptInstructions.pdf and ASPmanual2010.pdf 3.1.4 for more details
\author{Alice Allen,$^{1,10}$ G. Bruce Berriman,$^2$ Kimberly DuPrie,$^{3,1}$ Jessica Mink,$^4$ Robert Nemiroff,$^5$ Thomas Robitaille,$^6$ Judy Schmidt,$^1$ Lior Shamir,$^7$ Keith Shortridge,$^8$ Mark Taylor,$^9$ Peter Teuben,$^{10}$ and John Wallin$^{11}$}
\affil{$^1$Astrophysics Source Code Library, College Park, MD, US; \email{aallen@ascl.net}}
\affil{$^2$Infrared Processing and Analysis Center, California Institute of Technology, Pasadena, CA, US}
\affil{$^3$Space Telescope Science Institute, Baltimore, MD, US}
\affil{$^4$Smithsonian Astronomical Observatory, Cambridge, MA, US}
\affil{$^5$Michigan Technological University, Houghton, MI, US}
\affil{$^6$Freelance, Leeds, UK}
\affil{$^7$Lawrence Technological University, Southfield, MD, US}
\affil{$^8$Knave and Varlet, McMahons Point, NSW, Australia}
\affil{$^9$H.~H.~Wills Physics Laboratory, University of Bristol, U.K.}
\affil{$^{10}$Astronomy Department, University of Maryland, College Park, MD, US}
\affil{$^{11}$Middle Tennessee State University, Murfreesboro, TN, US}

% This section is for ADS Processing.  There must be one line per author.
\paperauthor{Alice Allen}{aallen@ascl.net}{}{Astrophysics Source Code Library}{}{College Park}{MD}{}{US}
\paperauthor{G. Bruce Berriman}{gbb@ipac.caltech.edu}{orcid.org/0000-0001-8388-534X}{California Institute of Technology}{IPAC}{Pasadena}{CA}{}{US}
\paperauthor{Kimberly DuPrie}{kduprie@stsci.edu}{}{Space Telescope Science Institute/ASCL}{}{Baltimore}{MD}{}{US}
\paperauthor{Jessica Mink}{jmink@cfa.harvard.edu}{orcid.org/0000-0003-3594-1823}{Smithsonian Astronomical Observatory}{}{Cambridge}{MA}{}{US}
\paperauthor{Robert Nemiroff}{nemiroff@mtu.edu}{orcid.org/0000-0002-4505-6599}{Michigan Technological University}{}{Houghton}{MI}{}{US}
\paperauthor{Thomas Robitaille}{thomas.p.robitaille@gmail.com}{orcid.org/0000-0002-8642-1329}{Freelance}{}{Leeds}{}{}{UK}
\paperauthor{Judy Schmidt}{gecko@geckzilla.com}{}{Astrophysics Source Code Library}{}{}{}{}{US}
\paperauthor{Lior Shamir}{lshamir@ltu.edu}{orcid.org/0000-0002-6207-1491}{Lawrence Technological University}{}{Southfield}{MI}{}{US}
\paperauthor{Keith Shortridge}{keithshortridge@gmail.com}{}{Knave and Varlet}{}{McMahons Point}{NSW}{}{Australia}
\paperauthor{M.~B.~Taylor}{m.b.taylor@bristol.ac.uk}{0000-0002-4209-1479}{University of Bristol}{School of Physics}{Bristol}{Bristol}{BS8 1TL}{U.K.}
\paperauthor{Peter Teuben}{teuben@astro.umd.edu}{orcid.org/0000-0003-1774-3436}{University of Maryland}{Astronomy Department}{College Park}{MD}{}{US}
\paperauthor{John Wallin}{John.Wallin@mtsu.edu}{orcid.org/0000-0001-5678-8325}{Middle Tennessee State University}{}{Murfreesboro}{TN}{}{US}

\begin{abstract}
The Astrophysics Source Code Library (ASCL) is a free online registry of research codes; it is indexed by ADS and Web of Science and has over 1300 code entries. Its entries are increasingly used to cite software; citations have been doubling each year since 2012 and every major astronomy journal accepts citations to the ASCL. Codes in the resource cover all aspects of astrophysics research and many programming languages are represented. In the past year, the ASCL added dashboards for users and administrators, started minting Digital Objective Identifiers (DOIs) for software it houses, and added metadata fields requested by users. This presentation covers the ASCL's growth in the past year and the opportunities afforded it as one of the few domain libraries for science research codes.
\end{abstract}

\section{Introduction}
The Astrophysics Source Code Library, started in 1999, is a free online registry of academic software used in research; as of this writing, it has nearly 1400 code entries. It is indexed by the SAO/NASA Astrophysics Data System (ADS) and Web of Science's Data Citation Index. This resource is increasingly used to cite software, with code entries cited in nearly 50 journals. All major astronomy journals and general science journals such as  \emph{Nature} and  \emph{Science} accept citations to ASCL entries. Figure \ref{ASCL_fig1} shows the main sources of ASCL citations.

\articlefigure[width=.47\textwidth]{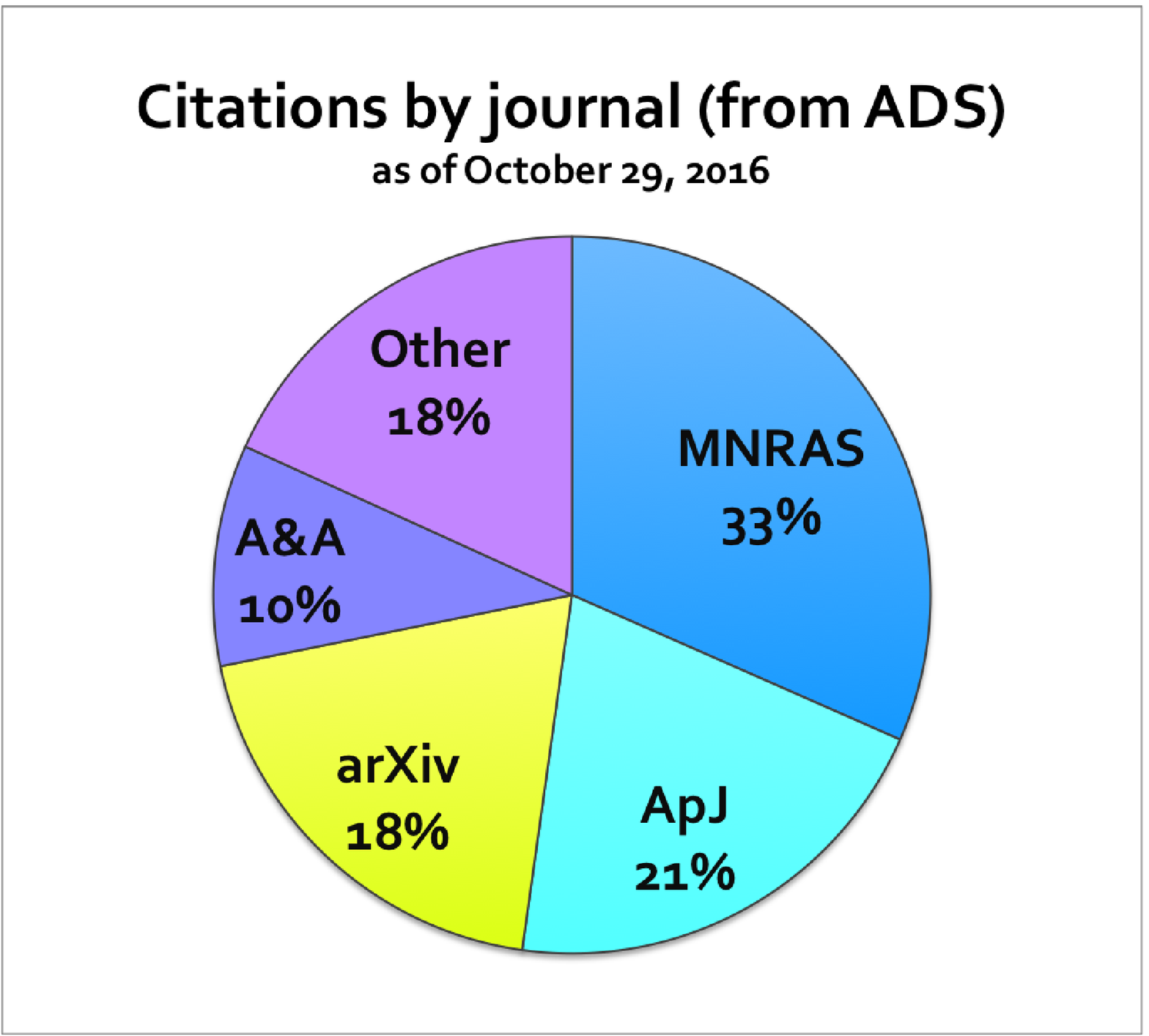}{ASCL_fig1}{Citations to ASCL entries by journal}
%\newline
%Data retrieved from \url{http://ascl.net/dashboard} October 29, 2016}

\section{Growth}
On average, eighteen codes a month are added to the ASCL, most of which are found by the editors, but increasingly, authors are submitting their codes for inclusion. Since 2014, 160 codes registered in the ASCL and assigned ASCL IDs were submitted by their authors; the rate of author submissions in 2016 is currently up 10\% over 2015. As figure \ref{ASCL_fig2} shows, citations to ASCL entries have doubled every year since 2012. 

\articlefigure[width=.45\textwidth]{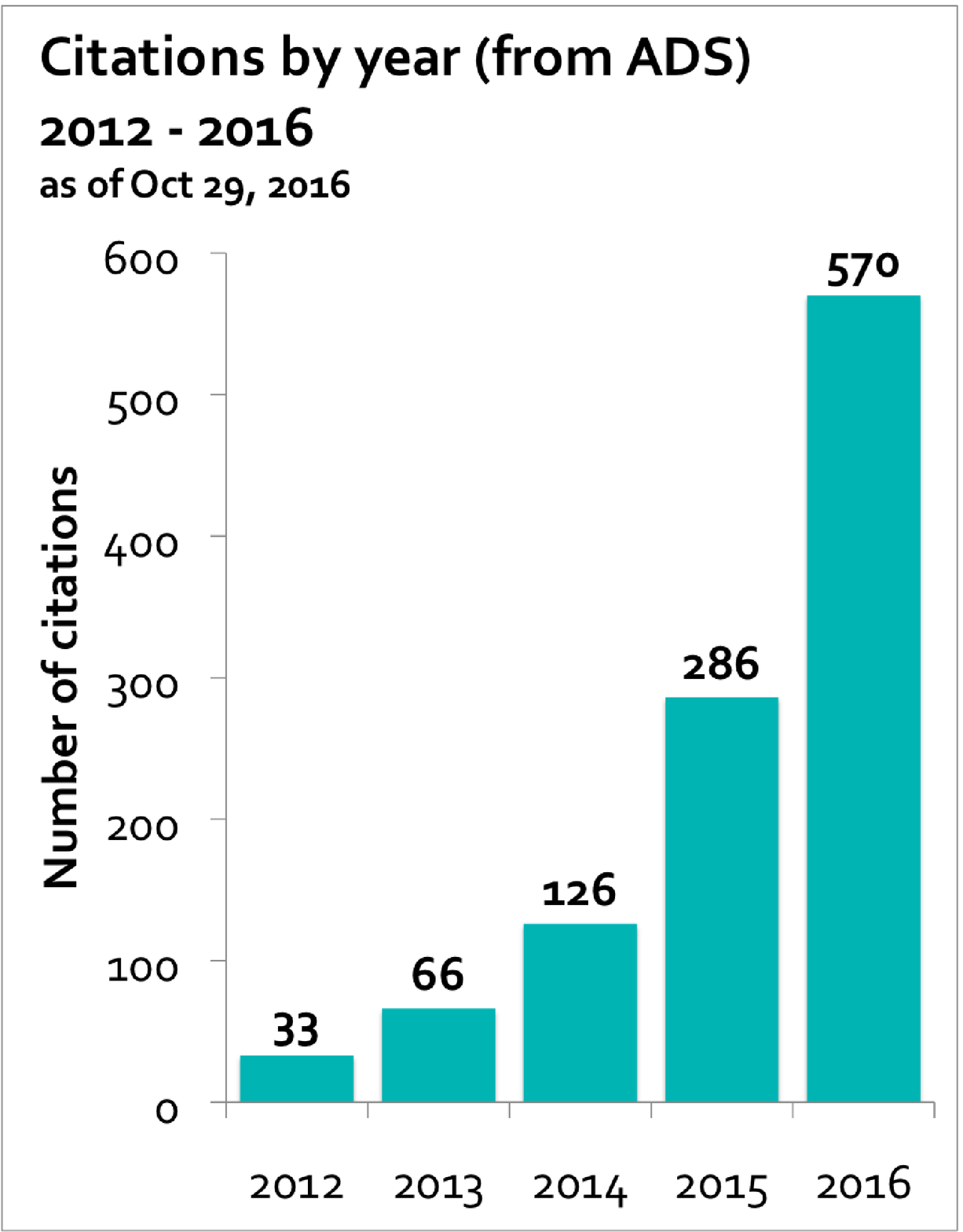}{ASCL_fig2}{Citations to ASCL entries by year}
%\newline
%Data retrieved from \url{http://ascl.net/dashboard} October 29, 2016}
%\articlefiguretwo{P8-1_f1.eps}{P8-1_f2.eps}{ASCL_fig1_fig2}{}

% \clearpage 

\section{New features added in 2016}
The ASCL can now mint Digital Objective Identifiers (DOIs) for codes it houses and serves for download through an agreement with University of Maryland Libraries, and all codes deposited with the ASCL have been assigned DOIs. The ASCL also now tracks DOIs for software it registers. We have added a field that allows code authors to list their preferred citation method in the code entry; other fields added (though at this time sparsely populated) include a category field and a "see also" field intended to alert readers to entries similar to the one being viewed. Dashboards for users and administrators have been added; the user dashboard shows:

\begin{itemize}

\item  number and percentage of site links working
\item  number of codes in the ASCL
\item  how many submitted entries are awaiting editor attention
\item  bar chart showing the number of codes added each year in the current and past four years
\item  bar chart showing the number of citations by year for the current and past four years
\item  pie chart showing citations by journal
\item  list of journals in which citations to ASCL appear
\item  list of the 10 most prolific software authors
\item  list of the 10 most viewed code entries
\end{itemize} 

In 2016, the Thomson-Reuters Web of Science Data Citation Index started indexing the ASCL. The service picks up ASCL entries quarterly through an automated report generated on demand. The ASCL also has been formalizing its relationships with other entities and since its last report at ADASS has registered with re3data.org, a DataCite registry of research data repositories.

\section{Influence and Advocacy}
As one of the few, and at this point, one of the more prominent, domain library for research codes, the ASCL is often asked to provide input and information on scientific software issues, and 2016 offered a number of opportunities to do so. 

ASCL editor Alice Allen was invited to present an enrichment session at the National Library of Medicine (US) in April 2016 for the National Digital Stewardship Residency (NDSR) program; the NDSR program is a project of the Library of Congress and the Institute of Museum and Library Services. Also in April, the ASCL participated in CodeMeta, a project "to create a crosswalk table between standards already in use \-- think of this as a Rosetta stone of software metadata."\footnote{CodeMeta project site, \url{http://codemeta.github.io/}} One working session of CodeMeta was conducted at the Force2016 conference;\footnote{Force2016, \url{https://www.force11.org/meetings/force2016}} in addition to the CodeMeta working session, the ASCL also participated in a workshop as a member of the Force11 Software Citation Working Group that was finalizing the now-published Software Citation Principles \citep{softwarecitationprinciples}.

Subsequently, Allen received an invitation to represent the ASCL at a Perspectives Workshop on Engineering Academic Software at Schloss Dagstuhl --- Leibniz Center for Informatics held in June 2016.\footnote{\url{http://www.dagstuhl.de/no_cache/en/program/calendar/semhp/?semnr=16252}} Thought most outputs from this workshop are still under development, a lightning talk on one, \emph{I solemnly pledge: A Manifesto for Personal Responsibility in the Engineering of Academic Software}\footnote{\url{http://wssspe.researchcomputing.org.uk/wp-content/uploads/2016/06/WSSSPE4_paper_15.pdf} (PDF)} was presented in September 2016 at the fourth Working towards Sustainable Software for Science: Practice and Experiences (WSSSPE) workshop held in Manchester, UK \citep{wssspe4}, which Allen also attended on behalf of the ASCL. Finally, the ASCL applied for and was granted attendance at the November 2016 OpenCon2016 meeting in Washington, DC.\footnote{\url{http://www.opencon2016.org/}}

\section{Conclusion}
The ASCL has continued its growth in citable entries and is seeing more use of its entries for citation with a doubling in the number of citations every year since 2012. It has increased services by minting DOIs for software it holds, creating a dashboard to provide information to users, and allowing authors to specify how they want their codes cited. The ASCL has become better known not just in astronomy, but as one of the few domain scientific code libraries, has also become known more generally; as a result, the ASCL is asked to participate in meetings and conferences that focus on open science and research software. 

\acknowledgements The ASCL is grateful for financial support from the Heidelberg Institute for Theoretical Studies (HITS), for facilities and services support from Michigan Technological University, the Astronomy Department at the University of Maryland, and the University of Maryland Libraries, and for travel support from CodeMeta and WSSSPE4.

\bibliography{P8-1} 

\begin{thebibliography}{}
\expandafter\ifx\csname natexlab\endcsname\relax\def\natexlab#1{#1}\fi
\expandafter\ifx\csname url\endcsname\relax
  \def\url#1{\texttt{#1}}\fi
\expandafter\ifx\csname urlprefix\endcsname\relax\def\urlprefix{URL }\fi
\providecommand{\eprint}[2][]{\url{#2}}

\bibitem[{Allen et~al.(2016)}]{wssspe4}
Allen, G., et~al. (eds.) 2016, Proceedings of the Fourth Workshop on
  Sustainable Software for Science: Practice and Experiences (WSSSPE4), no.
  1686 in CEUR Workshop Proceedings (Aachen).
  \urlprefix\url{http://ceur-ws.org/Vol-1686/}

\bibitem[{{Smith} et~al.(2016){Smith}, {Katz}, {Niemeyer}, \& {FORCE11 Software
  Citation Working Group}}]{softwarecitationprinciples}
{Smith}, A.~M., {Katz}, D.~S., {Niemeyer}, K.~D., \& {FORCE11 Software Citation
  Working Group} 2016, PeerJ Computer Science, 2:e86

\end{thebibliography}

\end{document}